\documentclass{article}

\usepackage[preprint]{nips_2018}

\usepackage[utf8]{inputenc} 
\usepackage[T1]{fontenc}    
\usepackage{hyperref}       
\usepackage{url}            
\usepackage{booktabs}       
\usepackage{amsfonts}       
\usepackage{nicefrac}       
\usepackage{graphicx}
\usepackage{subcaption}
\graphicspath{ {images/} }
\usepackage{microtype}      
\usepackage{amsmath}
\usepackage{comment}
\usepackage{natbib}
\DeclareMathOperator*{\argmin}

\title{Handling Cold-Start Collaborative Filtering with Reinforcement Learning}

\author{
  Hima Varsha Dureddy\thanks{Authors contributed equally} \\
  Language Technologies Institute \\
  School of Computer Science \\
  Carnegie Mellon University \\
  Pittsburgh, PA \\
  \And
  Zachary Kaden\footnotemark[1] \\
  Language Technologies Institute \\
  School of Computer Science \\
  Carnegie Mellon University \\
  Pittsburgh, PA \\
}

\begin{document}

\maketitle
\begin{abstract}

A major challenge in recommender systems is handling new users, whom are also called \textit{cold-start} users. In this paper, we propose a novel approach for learning an optimal series of questions with which to interview cold-start users for movie recommender systems. We propose learning interview questions using Deep Q Networks to create user profiles to make better recommendations to cold-start users. While our proposed system is trained using a movie recommender system, our Deep Q Network model should generalize across various types of recommender systems.

\end{abstract}

\section{Introduction}
There are a number of methods for recommender systems to suggest movies to users based on their likes, dislikes, similarities to other users, etc. Pure collaborative filtering methods base their recommendations on community preferences, ignoring user and item attributes, such as user demographics and product descriptions. Pure content-based filtering methods match query words or other user data with movie attribute information, ignoring data from other users. Additionally, several hybrid algorithms, such as \cite{schein2002methods}, elegantly combine collaborative filtering and content-based filtering methods.

One common challenge in these systems is handling cold-start users for whom we lack prior information. This occurs, for example, when providing recommendations to a new user. A popular method of handling cold-start situations is for recommender systems to initiate an interview, where users are asked a series of queries to build an initial user profile. Handling cold-start users with an interview was previously proposed in \cite{golbandi2011adaptive}, where the authors constructed a decision tree to conduct such an interview. In our work, we propose applying deep reinforcement learning to learn an optimal series of questions with which to interview cold-start users. Specifically, we propose a learner that, at each step in the interview, generates a query to maximize the information gain for a cold-start user based on his/her answers to previous queries. To the best of our knowledge, this is the first work that uses reinforcement learning to conduct such an interview.


The rest of this paper is organized as follows: Section 2 presents related works, Section 3 presents methodology, Section 4 presents experiments, evaluation methods, and results, Section 5 presents discussions, and Section 6 concludes.

\section{Related work}
A lot of recent work in collaborative filtering leverages reinforcement learning techniques. One approach by \cite{shani2005mdp} is to treat recommender systems as a \textit{Markov Decision Process} (MDP) and use RL techniques to solve it. \cite{leelreinforcement} also formalize the problem as a MDP and learn the connection between the time sequence of user ratings using Q-learning. \cite{choi2018reinforcement} formulate the problem as a gridworld game using a a bi-clustering technique to reduce the action and state space substantially. To address the cold-start problem using RL techniques, \cite{nguyen2014cold} cast it as a contextual bandit problem. They propose a new method based on the popular LinUCB algorithm \cite{li2010contextual} for contextual-bandit problems.

In the past, \cite{schein2002methods} combined content and collaborative data under a single probabilistic framework. \cite{park2009pairwise}  proposed a pairwise predictive system to build feature-based regression models, where user demographic information, item content features, and other information about users and items are utilized to address cold-start problems. The model uses a user's ratings as targets and uses a pair of vectors $x_u$ and $z_i$, to predict the rating $r_{ui}$ on the item $z_i$ given by the user $x_u$, where $u$ is an index over users and $i$ is an index over items.

The indicator function is represented as:
\begin{equation}
	s_{ui} = x_u W z_i^T = w^T (z_i \otimes x_u)
\end{equation}

where $(z_i \otimes x_u)$ represents the outer product of $z_i$ and $x_u$. The optimization equation is as follows:
\begin{equation}
	\arg\min_{w} \sum_{ui \in O} (r_{ui} - s_{ui})^2 + \lambda ||w||_2^2
\end{equation}

A statistical approach for collaborative filtering is Probabilistic Matrix Factorization (PMF), \cite{mnih2008probabilistic}. PMF scales linearly with the number of observations and performs well on sparse data. For example, assume a ratings matrix $R$ where each row corresponds with a user and each column corresponds with a movie. If $R$ has dimension $N\times M$, PMF extracts an $N\times D$ matrix $U^T$ and a $D \times M$ matrix $V$ such that $R=U^TV$ where D is the dimension of the latent space. An extension of this is the Bayesian treatment of PMF called as Bayesian Probabilistic Matrix Factorization (BPMF), \cite{salakhutdinov2008bayesian}, which has been shown to outperform simple PMF.

One successful method for addressing cold-start is a preliminary interview process to learn user's interests for good recommendation. \cite{golbandi2011adaptive} propose building a decision tree for the initial interview process by bootstrapping. They construct a decision tree for the initial interview with each node being an interview question, enabling the recommender to query a user adaptively according to his/her prior responses.

In \cite{zhou2011functional}, the authors build on the previous work and propose \textit{functional matrix factorization}. In this, latent profiles are associated for each node of the decision tree and hence the user profile is the function of all possible answers to the interview. The novelty is an iterative optimization algorithm that alternates between decision tree construction and latent profiles' generation. This helps to learn the best decision tree to ask questions and generate good embeddings simultaneously.
The profile $u_i$ is tied to user $i$’s responses in the form of a function; thus the name functional matrix factorization (fMF). Given an answer set $a_i$, the user profile is generated by $u_i = T(a_i)$. The goal is to learn T and $v_j$ (item profiles) from the observed ratings set.

\begin{equation}
	T, V = \arg\min_{T \in H, V} \sum_{(i, j) \in O} (r_{ij} - v_j^T T(a_i))^2 + \lambda ||V||^2
\end{equation}

where the user profile $u_i = T(a_i)$ is modeled as a function of the answers a user gives based on the interview questions. $r_{ij}$ represents the actual item rating and $v_j$ is the latent representation of the item profile.

\section{Methodology}

We propose a novel interview based method to address the cold-start problem in collaborative filtering. To perform collaborative filtering, BPMF, \cite{salakhutdinov2008bayesian}, is used to create movie embeddings based on the ratings from warm start users that already exist in the dataset. To generate user embeddings for cold-start users, we propose an interview method. For asking the ideal questions, we implement a Deep Q Network that generates interview questions to be answered by a cold-start user. The interview consists of the DQN generating questions of the form: ``Do you like $\lambda?",$ where lambda is drawn from the set of all movies in the action space. The user may respond to the interview question by rating the movie 1-5, or by responding that he or she has not seen the movie, for which we consider the rating to be 0. Based on a user's response to the previous question(s), subsequent questions are dynamically generated by the DQN. Once the interview has been completed, the question answer pairs are passed into a multilayer perceptron model that generates a predicted user embedding. Once we have a user embedding and a movie embedding, we can model a movie rating
$$
\hat{r}_{ij} = f(u_{i}, m_{j})
$$
for some function $f,$ where $u_{i}$ is the user embedding for user $i$, $m_{j}$ is the item embedding for movie $j$, and $\hat{r}_{ij}$ is the predicted rating for movie $j$ by user $i$.

We train our Deep Q Network such that the state is a representation of the set of question-answer pairs derived from the interview. After each interview question, the state is updated and passed through the DQN to generate the next question. The action space of the DQN is the set of all movies about which an interview question may be generated. Once we have completed the interview, the terminal state is passed into the MLP to create a predicted user embedding. The predicted user embedding is then used to generate predicted ratings for all movies by the cold-start user.

To train our DQN, we calculate the RMSE of the movie rating predictions and consider the inverse of the RMSE to be the reward for our actions. Given an interview of length $k,$ and corresponding actions $a_{1}, \cdots, a_{k}$ we receive an immediate reward of 0 for all actions $a_t, t<k.$ The reward $r_{k}$ for action $a_{k}$ is the inverse RMSE of the predicted movie ratings. That is,
$$
r_{k} = \frac{1}{\text{RMSE}(\hat{M}_{i\{j\}}, M_{i\{j\}})},
$$
where $\hat{M}_{i\{j\}}$ is the set of the predicted movie ratings for user $i$ and $M_{i\{j\}}$ are the actual movie ratings for user $i$. Thus, the reward for an action $a_{t}$ is
$$
r_{t} = \frac{\gamma^{k-t}}{\text{RMSE}(\hat{M}_{i\{j\}}, M_{i\{j\}})},
$$
where gamma is the discount rate.

In the typical Q-Learning setting, we only update the Q function for the observed actions we take. We perform a slightly modified update to our Q-function where we additionally set the reward for any previously selected action in a state to be 0. That is, we update the Q-function such that the q-value for repeating an interview question is 0.

The MLP is trained using supervised learning. We explore training the MLP in two different manners. In the first setting, the MLP is trained with the user embedding generated from the BPMF model as the true value. In the second setting, the MLP is directly trained on a user's actual movie ratings. The input to the former MLP is the terminal state of the DQN and the input to the latter MLP is the terminal state of the DQN, along with the movie ID whose rating is to be predicted and a movie rating constant that is the mean movie rating in the dataset. The output of the MLP differs in the two settings. The MLP trained on user embeddings outputs a predicted user embedding, whereas the MLP trained on movie ratings outputs a predicted rating for a specific movie. For ease of exposition, we refer to the first model as our Q-Embedding model and we refer to the second model as our Q-Rating model.

\begin{figure}[h]
\centering
\includegraphics[width=13cm]{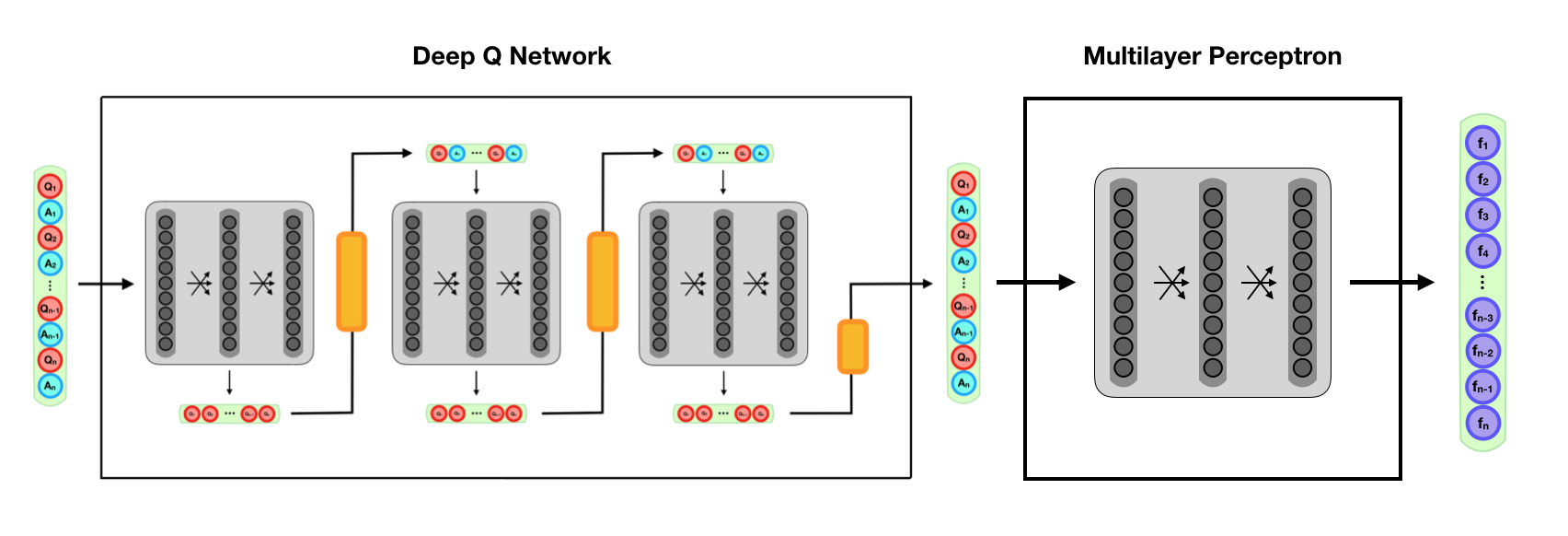}
\caption{The Q-Embedding Model}
\label{fig: qembedding}
\end{figure}

\subsection{Q-Embedding Model}
The Q-Embedding model is pictured in Figure \ref{fig: qembedding}. The model is a composite network that combines a DQN and an MLP. The DQN takes as input an initial state vector of dimension twice the action space. This dimension is needed to represent each possible interview question-answer pair. The DQN outputs a vector of q-values of dimension equal to the action space. Given the DQN output, we select an interview question to ask, and update the state depending on the user's answer. The new state is then passed back into the DQN until we have completed our interview. After the final question is asked, we update the state a final time, and pass the terminal state into the MLP. The MLP learns a function to transform the terminal state into a user embedding. The DQN is trained using the reward function described above, and the MLP is trained using the BPMF user embeddings as the true value.

We implement the Q-Embedding model for an interview of length 3. The DQN action space consists of the 100 most frequently rated movies in the data set. The DQN has two hidden layers. It takes as input a state vector of dimension 200. The first hidden layer is a dense layer with 64 neurons and a relu activation. The second hidden layer is a dense layer with 32 neurons and a relu activation. At each layer we utilize a dropout rate of 0.5 to regularize and prevent overfitting. The output layer is dense layer with 100 neurons and a relu activation. We use the Adam optimizer with a learning rate of 5e-4 and categorical cross-entropy loss function.

The MLP takes as input a 200 dimensional state vector. The MLP has two hidden layers. Both hidden layers are densely connected layers with 32 neurons and relu activation functions. Both layers utilize a dropout rate of 0.5 to regularize and prevent overfitting. The output layer is a dense layer with 10 neurons and a tanh activation. The MLP uses the Adam optimizer with a learning rate of 1e-4 and mean squared error loss function.

\begin{figure}[h]
\centering
\includegraphics[width=13cm]{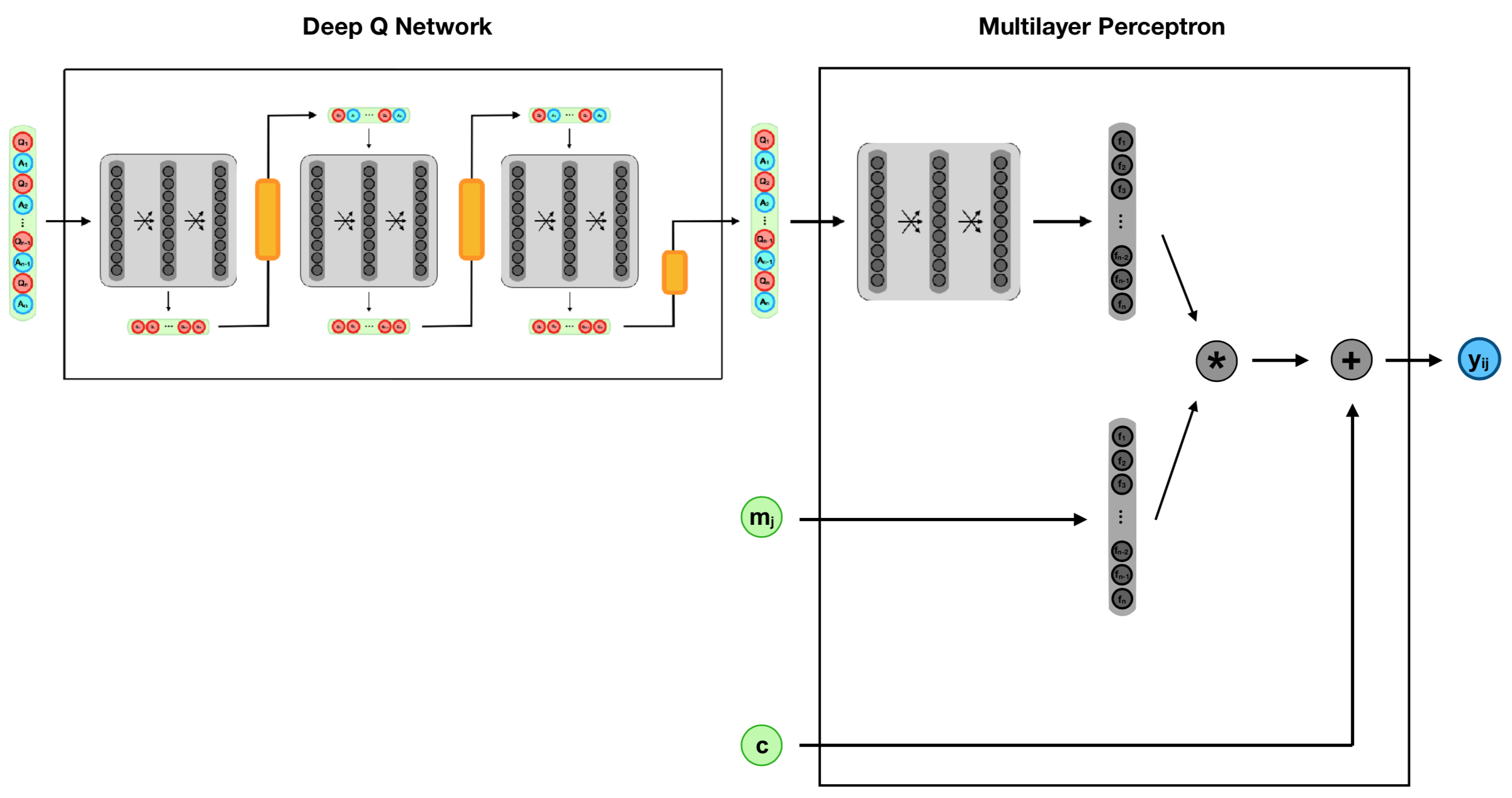}
\caption{The Q-Rating Model}
\label{fig: qrating}
\end{figure}

\subsection{Q-Rating Model}
The Q-Rating model is pictured in Figure \ref{fig: qrating}. The model is also a composite network that combines a DQN with an MLP. The DQN behaves similarly to the DQN in the Q-Embedding model. The MLP for the Q-Rating model is slightly different than the MLP for the Q-Embedding Model. Unlike the Q-Embedding model, the Q-Rating model directly predicts the user rating for a movie. The Q-Rating MLP takes three inputs: the terminal state from the DQN, a movie id, and the average movie rating. Note that the average movie rating is constant across all movies and terminal states. The MLP embeds the user and movie into latent spaces. The MLP creates a user embedding using the architecture as in the Q-Embedding model. Additionally, the MLP embeds the movie into a latent item space. The movie embedding is pre-trained with the BPMF movie embeddings; however, the embedding layer is retrained in parallel with the user embeddings. After the MLP embeds the user and movie it takes the inner product of the two embeddings, and adds the mean movie rating to that product. The sum of the mean movie rating and the inner product of the embeddings is the output as the user movie rating prediction.

We implement the Q-Rating model for an interview of length 3. The DQN action space consists of the 100 most frequently rated movies in the data set. The DQN has two hidden layers. It takes as input a state vector of dimension 200. The first hidden layer is a dense layer with 64 neurons and a tanh activation. The second hidden layer is a dense layer with 32 neurons and a tanh activation. At each layer we utilize a dropout rate of 0.5 to regularize and prevent overfitting. The output layer is dense layer with 100 neurons and a relu activation. We use the Adam optimizer with a learning rate of 5e-4 and categorical cross-entropy loss function.

The MLP for the Q-Rating model features three streams. The first stream takes a state vector as input and creates a user embedding with the same architecture as the the MLP in the Q-Embedding model. The second stream takes as input a movie. This stream has a single embedding layer that embeds the movie into a 10 dimensional latent space. The first two streams merge together by taking the inner product between them. Thus, after the merge only a single scalar value remains. The third stream takes as input the mean movie rating in the dataset. This third stream is added to the inner product of the first two streams. The resulting sum is the output of the MLP, a one dimensional scalar representing the predicted user movie rating. The MLP is trained using the Adam optimizer with a learning rate of 1e-4 and a custom loss function that clips the predicted ratings to be between 1 and 5 before computing the mean squared error of the true ratings and the predicted ratings.

\subsection{Hyperparameters}
Both models are trained using a discount rate of $\gamma = 1.$ While training the models, we select actions using an $\epsilon$-greedy policy. We begin training with $\epsilon=1$ and decrement $\epsilon$ by 5e-2 per epoch until $\epsilon=0.2.$ We train the models utilizing retraining. That is, after the models appear to stabilize, we restart training using the best performing model. When restarting training, we reset $\epsilon=1$ and lower the DQN learning rate to 1e-5.

\section{Experiments}
\subsection{Datasets}
We utilize the \cite{movielens}, dataset to train and test our proposed model. The MovieLens dataset contains one million ratings along with movie names. Information about this datasets is provided in Table \ref{db_stats}.

\begin{table}[t]
 \begin{center}
  \begin{tabular}{|c|c|c|c|c|c|}
  	\hline
    Dataset     & No. of Users     & No. of Movies & No. of Ratings \\
    \hline
    MovieLens 1M & 6040  & 3706 & 1,000,209 \\
    \hline
  \end{tabular}
  \end{center}
  \caption{Statistics of the MovieLens dataset, \cite{movielens}}
  \label{db_stats}
\end{table}

\subsection{Evaluation}

To evaluate the proposed model, we calculate the RMSE of the predicted user ratings and the actual user ratings. To do this, we divide the users into a 75/25 split for training and testing, respectively. As is done in \cite{zhou2011functional}, for testing we divide the movies using a 75/25 split, where 75\% of movies comprise the interview set and 25\% comprise the test set. The interview set consists of the possible answers to interview questions. If a query is asked about a movie in the test set, then it is treated as an unwatched movie by the user, even if the user has rated the movie.

We evaluate our results against the following models:
\begin{itemize}
\item Tree, \cite{golbandi2011adaptive}: This model builds a decision tree without using latent user/item profiles. It fits the tree based on the ratings from the dataset.

\item TreeU, \cite{golbandi2011adaptive}: This model uses matrix factorization to extract the movie and user profiles. Based on that, a decision tree is constructed to fit the latent profiles. User profiles from the decision tree are used with movie profiles from the matrix factorization.

\item Functional Matrix Factorization (fMF), \cite{zhou2011functional}: This model constructs a decision tree. In addition, the model implements an iterative optimization algorithm that alternates between decision tree construction and latent profile extraction.


\end{itemize}

\subsection{Results}
We learn a 3-question cold-start interview. The movie-action space is a set of the 100 most frequently rated movies. Additionally, we also use our learned policy to conduct a a 4-question cold-start interview. We examine results of the learned policies for the Q-Embedding model and the Q-Rating model.
\begin{figure}[h]
\centering
\includegraphics[scale=0.45]{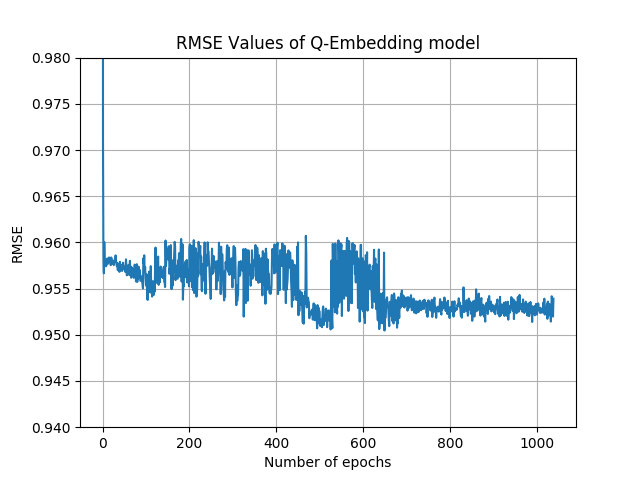}
\caption{Testing RMSE for Q-Embedding model}
\label{fig: qembedding rmse}
\end{figure}

Figure \ref{fig: qembedding rmse} shows the RMSE values recorded for the users in the test set per epoch during the training phase for the Q-Embedding model. It can be observed that the Q-Embedding model hits the minimum RMSE of 0.9507 at around 520 epochs.
\begin{figure}[h]
\centering
\includegraphics[scale=0.45]{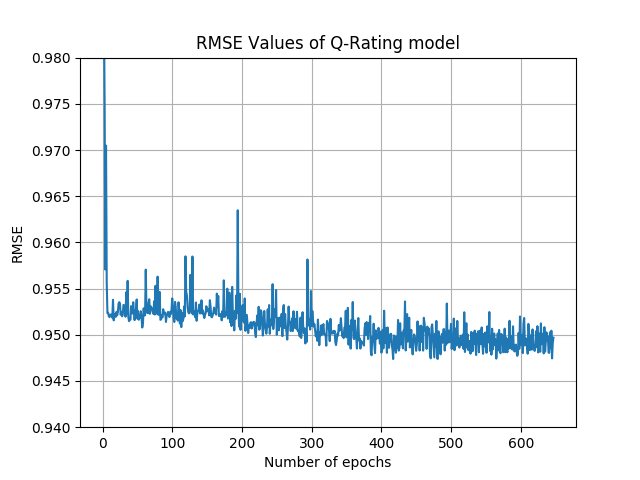}
\caption{Testing RMSE for Q-Rating model}
\label{fig: qrating rmse}
\end{figure}

Figure \ref{fig: qrating rmse} shows the RMSE values recorded for the users in the test set per epoch during the training phase for Q-Rating model. The Q-Rating model hits the minimum RMSE of 0.9472 at around 480 epochs.
\begin{table}[t]
  \begin{center}
  \begin{tabular}{|c|c|c|}
  	\hline
    Model & 3 questions & 4 questions \\
    \hline
    \hline
    Tree & 0.9767 & 0.9683\\
    \hline
    TreeU &  0.9913 & 0.9887\\
    \hline
    fMF & 0.9509 & 0.9480\\
    \hline
    Q-Embedding model & \textbf{0.9507} & \textbf{0.9486}\\
    \hline
    Q-Rating model & \textbf{0.9472} & \textbf{0.9469} \\
    \hline
  \end{tabular}
  \end{center}
  \caption{Movie rating RMSE values on the test set}
  \label{results}
\end{table}
\\

Table \ref{results} compares our results with other existing approaches: Tree, TreeU and fMF. We observe that both our models outperform Tree, TreeU, and fMF models' results for a 3-question interview, with the Q-Rating model performing the best. Additionally, our Q-Rating model outperforms all other models for a 4-question interview. This is particularly impressive given that the model was only trained to conduct 3-question interviews.

\section{Discussion}
\begin{table}
\begin{center}
\label{interview1}
\begin{tabular}{| l | l | l | l |}
\hline
Turn & Movie              & Genre                    & Rating \\
\hline
\hline
1    & The Usual Suspects & Crime, Mystery, Thriller & 5      \\
\hline
2    & The Matrix         & Action, Sci-Fi           & 2      \\
\hline
3    & Clerks             & Comedy                   & 0   \\
\hline
\end{tabular}
\vspace*{5pt}
\centering
\label{interview2}
\begin{tabular}{| l | l | l | l | }
\hline
Turn & Movie               & Genre                      & Rating \\
\hline
\hline
1    & The Usual Suspects  & Crime, Mystery, Thriller   & 0      \\
\hline
2    & Alien               & Action, Sci-Fi             & 2      \\
\hline
3    & Romancing the Stone & Adventure, Comedy, Romance & 0     \\
\hline
\end{tabular}
\end{center}
\caption{Sample interviews}
\label{interviews}
\end{table}

Table \ref{interviews} presents sample interviews for two users. The first interview question (which is constant for all users) is \textit{The Usual Suspects}. Based on the user's response, the next interview question is selected. An interesting observation from the interview samples is that each interview question asks the user about a movie from a different genre. Given that no movie meta-information was made available to the learner, this evidences particularly strong reasoning abilities by the Q-Network over the movie space. Intuitively, one expects that asking questions about a diverse set of movies helps the model learn holistically about the user.

Comparing the Q-Rating and the Q-Embedding models from Figures \ref{fig: qembedding rmse} and \ref{fig: qrating rmse}, we observe that the Q-Rating model is more stable than the Q-Embedding model. The Q-Embedding model policy diverges after about 500 epochs, before relearning a successful policy. The Q-Rating model, on the other hand, doesn't diverge and more steadily improves or maintains its learned policy.

We also observe that the Q-Rating model outperforms the Q-Embedding model. This behavior is consistent with expectations, as the Q-Rating model learns the movie embeddings in parallel with the user embeddings, whereas the Q-Embedding model utilizes static BPMF movie embeddings. Further, the Q-Rating model trains by directly minimizing the loss on the RMSE of the predicted movie ratings, while the Q-Embedding model indirectly minimizes the predicted movie ratings RMSE by minimizing the MSE of the predicted user embedding and the BPMF user embedding.

The Q-Embedding model takes approximately 1.3 seconds per one hundred users to train on a MacBook Pro with a 2.7 GHz Core i5 processor and 8GB of RAM. The Q-Rating model takes approximately 3.2 seconds per one hundred users training on the same device. Hence, the Q-Rating model takes approximately 3x longer to train on the MovieLens dataset. Given this, it may be the case that the Q-Embedding model is preferred when working with extremely large datasets (e.g. the Netflix dataset of 100 million ratings). We note that the per epoch train time for the Q-Embedding model is independent of the size of the dataset. In contrast, the Q-Rating model may not scale to large datasets.

Impressively, despite the models being trained for 3-question interviews, both models generalize well to 4-question interviews. The Q-Embedding model decreases its RMSE by 2e-3 by asking a fourth question and the Q-Rating model decreases its RMSE by 3e-4. We posit that these RMSE decreases would be even larger if the models were trained to conduct a 4-question interview.

In addition to the improved performance offered by our models, we believe that our model architecture offers a number of innate advantages over the decision tree based models.

One such advantage of estimating the Q function with reinforcement learning is that the cold-start interview can consist of an arbitrary number of questions. This allows more patient users to spend more time tuning their profiles under the cold-start setting, and less patient users to shorten their interviews. Another advantage of our models is that they can be applied to warm-start users. Given a set of movie ratings from a warm-start user, we can create a state input for our DQN and generate new interview questions. This can be used to refine a warm-start user's embedding. In contrast, a decision tree based interview would not be able to generate ideal interview questions for warm start users. Given a warm-start user and a decision tree, the interview would need to begin at an arbitrary node. If the starting node were too deep into the tree, the interview would fail to ask questions from nodes occurring at lower depths in the tree. Conversely, if the starting node were too shallow in the tree, the interview would fail to consider the full context of a warm start user's ratings. A third advantage of our model is that the estimated q-values from the DQN could be used for active learning. The estimated q-values provide a measure of the importance of asking a user a given questions at a point in time. We envision two potential applications for this. One application is that the model could dynamically prompt an interview for a warm start user if it estimated the marginal information gain for asking a question was significant. The other application is that the model could dynamically stop an interview if it estimated a small marginal gain for asking additional questions.

\section{Further Work}
Further research could include improving the DQN's performance over large action spaces. Our DQN performs well when learning over the 100 most frequently rated movies; however, model performance degrades when the action space is increased. Ideally, a system should be able to handle an action space of all movies. To improve performance over a larger action space, we implemented Dueling Deep Q Networks, \cite{wang2015dueling}, as well as policy gradient methods, such as REINFORCE, \cite{williams1992simple}. Both of these methods have been known to improve performance over large action spaces. Unfortunately, neither of these methods were able to solve a large action space as well as a DQN could solve the restricted action space. One possible method for navigating a large movie action space would be to develop a composite Q network that uses multiple Q learners to first select the movie genre to ask, and then subsequently select the appropriate movie to query on within the chosen genre. Another avenue of further research could include incorporating movie and user metadata.

\section{Conclusion}

We've developed a composite Deep Q Network to learn an interview to handle cold-start users in movie recommender systems. To the best of our knowledge, our models outperform all other comparable methods. Additionally, our models possess several inherit advantages in comparison with decision tree based methods, such as accommodating warm start users and performing active learning. While we train our models in movie recommender setting, our proposed method should generalize broadly across all types of recommender systems.

\bibliography{ref}
\bibliographystyle{plainnat}

\end{document}